\documentclass[a4paper,11pt]{article}

\usepackage[dvips]{graphicx,psfrag}
\usepackage{amssymb}

\makeatletter

 \@addtoreset{equation}{section}
\makeatother
\newcommand{\beq}{\begin{eqnarray}}
\newcommand{\eeq}{\end{eqnarray}}
\def\tr{\mathop{\mathrm{tr}}\nolimits}

\begin{document}

\begin{titlepage}

\begin{flushright}
OU-HET 410\\
{\tt hep-th/0205036}\\
May 2002
\end{flushright}
\bigskip

\begin{center}
{\LARGE\bf
Wilsonian Renormalization Group Approach to ${\cal N}=2$ Supersymmetric Sigma Models
}
\vspace{1cm}

\setcounter{footnote}{0}
\bigskip

\bigskip
{\renewcommand{\thefootnote}{\fnsymbol{footnote}}
{\large\bf Kiyoshi Higashijima\footnote{
     E-mail: {\tt higashij@phys.sci.osaka-u.ac.jp}} and
 Etsuko Itou\footnote{
     E-mail: {\tt itou@het.phys.sci.osaka-u.ac.jp}}
}}

\vspace{4mm}

{\sl
Department of Physics,
Graduate School of Science, Osaka University,\\ 
Toyonaka, Osaka 560-0043, Japan \\
}

\end{center}
\bigskip

\begin{abstract}

We derive the Wilsonian renormalization group equation in two dimensional ${\cal N}=2$ supersymmetric nonlinear sigma models.  
This equation shows that the sigma models on compact Einstein K\"{a}hler manifolds are aymptotically free.
This result is gerenal and does not depend on the specific forms of the K\"{a}hler potentials.
We also examine the renormalization group flow in a new model which connects two manifolds with different global symmetries.

\end{abstract}

\end{titlepage}

\pagestyle{plain}

\section{Introduction}
Possible interactions in field theories are tightly restricted by the requirement of renormalizability. In four-dimensional scalar field theories, for example, only $\phi^3$ and $\phi^4$ interactions are renormalizable at least in perturbation theories. In two dimensions, however, nonpolynomial interactions like $\sin{\phi}$ or four fermion interactions are also renormalizable, and various interesting phenomena such as boson-fermion correspondence or dynamical mass generation are studied without ambiguities due to renormalization.  Among field theories with nonpolynomial interactions, nonlinear sigma models provide particularly interesting models in which field variables take values on curved manifolds (target space) like the sphere. Known examples on compact manifolds such as $S^N,\ CP^N$ or $Q^N$ are asymptotically free and generate the dynamical mass. Some of these theories curiously produce the diquark scalars, gauge fields and fermions as bound states  \cite{HKNT}. 

Recently, the supersymmetric nonlinear sigma models (SNL$\sigma$M) on Ricci-flat K\"ahler manifolds are explicitly constructed  \cite{HKN}. These theories have ${\cal N}=2$ supersymmetry, and are supposed to describe conformally invariant field theories. Although these field theories are interesting for their own sakes, they are also important as possible candidates for the sigma models description of the superstrings propagating in the curved background. Therefore it is curious to study the renormalization properties of ${\cal N}=2$ SNL$\sigma$M. In perturbation theories, $\beta$ function is proportianal to the Ricci tensor of the target space in the one loop order \cite{Morozov}.

Field theories have rich structure in two dimensions since any polynomial interactions are renormalizable in perturbation theories. Relations among these infinite number of field theories have been studied by using the so-called exact renormalization group equation, which is the Wilsonian renomalization group (WRG) method under the continuous change of the cutoff \cite{Wilson Kogut} \cite{Wegner and Houghton}. In practice, we have to introduce some kind of truncation of the functional differential equation describing the flow of an infinite number of coupling constants. The relevant truncation in the infrared region relies on the derivative expansion of the effective action \cite{Morris}. The simplest is the local potential approximation, in which one maintains only the potential term without any derivative \cite{Aoki}.

In this paper, we derive the Wilsonian renormalization group equation for the in the first nontrivial order of the derivative expansion in ${\cal N}=2$ SNL$\sigma$M. Because of the reparametrization invariance of the target manifold, the Lagrangian of the nonlinear sigma model is proportial of the metric of the target manifold and the potential term is absent. Therefore, the first nontrivial order of the derivative expansion is already of the second order in the derivatives \cite{CHL} \cite{CL}.

In this order, we obtain the $\beta$ function for the sigma model
\beq
\beta(g_{i \bar{j}})=\frac{1}{2\pi} R_{i \bar{j}}+\gamma \Big[\varphi^k g_{i \bar{j},k}+\varphi^{*\bar{k}}g_{i \bar{j},\bar{k}}+2g_{i \bar{j}} \Big].
\eeq
This $\beta$ function has the one loop correction term and the field rescaling term.
We also give the $\beta$ functions in case of D=3 ${\cal N}=2$ and D=4 ${\cal N}=1$ SNL$\sigma$M.

Using this $\beta$ function, we will show how the target spaces are deformed according to nonperturbative renormalization group flow.
We will show, in particular, the theories are asymptotically free when the target spaces are generic compact Einstein K\"{a}hler manifolds.
We will also construct a new model.
The model has two parameters, whose renormalization group flow interpolates two Einstein K\"{a}hler manifolds with different global symmetry.

This paper is organized as follows.
In section 2, we review the WRG for the most generic action and the approximation method based on the derivative expansion.
In section 3, we derive the WRG equation for two dimensional SNL$\sigma$M.
We also give the explicit WRG equation for the SNL$\sigma$M in three and four dimensions.
In two dimensions, we show the sigma medel with compact Einstein K\"{a}hler manifolds as target spaces are asymptotically free in section 4.
In section 5, we construct a new model.
This model has two parameters in the theory, whose WRG flow interpolates two Einstein K\"{a}hler manifolds with different global symmetries.

\section{Wilsonian Renormalization Group (WRG): Review}
Consider a general Euclidean quantum field theory for the fields $\Omega_i$ in D dimensions. The Euclidean path integral is
\beq
Z=\int [D \Omega_i] \exp \left[-S[\Omega] \right].
\eeq
Here $S[\Omega]$ is the most generic Euclidean action, which has the form
\beq
S[\Omega] \equiv \sum_n \frac{1}{n!} \int_{p_1} \cdots \int_{p_n} \hat{\delta}^{(D)}(p_1+\cdots +p_n) g_{i_1, \cdots ,i_n}(p_1,\cdots,p_n)\Omega_{i_1}(p_1) \cdots \Omega_{i_n} (p_n),\nonumber\\
\eeq
where $g$ is a coupling constant, and 
\beq
\hat{\delta}^{(D)} \equiv (2 \pi)^D \delta^{(D)}
\eeq
is the D dimensional delta-function.
The Fourier transformation of $\Omega$ is defined by
\beq
\Omega(x) \equiv \int_p \Omega(p) e^{-ipx}
\eeq
in which
\beq
\int_p \equiv \int \frac{d^D p}{(2 \pi)^D}.
\eeq

The basic idea to define the effective action is as follows.
We divide all fields $\Omega_i$ into two group: the high frequency modes ($\Omega_{i >}$) and the low frequency modes ($\Omega_{i <}$). After the higher modes are integrated out, the Wilsonian effective action ($S_{eff}$) is obtained:
\beq
Z&=&\int [D \Omega_i] \exp \left[-S[\Omega_i] \right]\nonumber\\
&=&\int[D \Omega_{i >}] [D \Omega_{i <}] \exp \left[-S[\Omega_{i <},\Omega_{i >}] \right]\nonumber\\
&\equiv&\int[D\Omega_{i <}] \exp \left[-S_{eff}[\Omega_{i <}] \right].
\eeq
Our aim is to find the flow of $S_{eff}$ by continuously changing the cutoff $\Lambda$ which is the boundary between higher and lower momentum.
For that purpose, we change the cutoff $\Lambda$ infinitesimally to $\Lambda(\delta t)=\Lambda e^{-\delta t}$. 
Thus in the effective action at $\Lambda$, we divide all fields into the higher modes[$\Omega_s$] and the lower modes again. 
The fields [$\Omega_S$] which have nonzero value only for the shell momentum between $\Lambda(\delta t)$ and $\Lambda$, are integrated out, and we can obtain the effective action for new cutoff $\Lambda(\delta t)$.
Hereafter, we simply write $S$ for $S_{eff}$, and we assume that Z is cutoff independent.
\beq
Z&=&\int [D \Omega]_{\Lambda(\delta t)} [D \Omega_s] \exp \left[-S[\Omega +\Omega_s ;\Lambda] \right]\nonumber\\
&=&\int [D \Omega]_{\Lambda(\delta t)} [D \Omega_s] \exp \Bigg[ - \left( S[\Omega; \Lambda]| +\frac{\delta S}{\delta \Omega_i} \Omega_s^i +\frac{1}{2}\Omega_s^i \frac{\delta^2 S}{\delta \Omega^i \delta \Omega^j}\Omega^j_s +O(\Omega_s^3) \right) \Bigg]\nonumber\\
&=&\int[D\Omega]_{\Lambda(\delta t)} \exp \Bigg[ -\left(S[\Omega; \Lambda]| +\frac{1}{2} \int_{p'} \tr \ln \left(\frac{\delta^2 S}{\delta \Omega^i \delta \Omega^j} \right) \right. \nonumber\\
&&\left. -\frac{1}{2} \int_{p'} \int_{q'} \frac{\delta S}{\delta \Omega^i} \left(\frac{\delta^2 S}{\delta \Omega^i \delta \Omega ^j} \right)^{-1} \frac{\delta S}{\delta \Omega^j}+O((\delta t)^2) \right) \Bigg] \label{expan1} \\
&\equiv&\int[D\Omega]_{\Lambda(\delta t)} \exp \left[-S[\Omega;\Lambda(\delta t)] \right],\label{expan2}
\eeq
where 
\beq
\int_{p'}&=&\int \frac{d \Omega_D}{(2\pi)^{D}} \int^{\Lambda}_{\Lambda(\delta t)} dp' \sim O(\delta t),
\eeq
with $\int d\Omega_D$ being the surface integral of the D-dimensional unit sphere.
We can drop terms of order $O(\Omega_s^3)$, because such terms vanish when $\delta t \rightarrow 0$.
In Eq.(\ref{expan1}) the action $S[\Omega; \Lambda]|$ is obtained from $S[\Omega; \Lambda]$ by dropping all fields with momentum above $\Lambda(\delta t)$
\beq
S[\Omega; \Lambda]|=\sum_n \frac{1}{n!} \int_{p_1} \cdots \int_{p_n}\hat{\delta}(p_1+\cdots +p_n) g(\Lambda) \Omega_{i_1} \cdots \Omega_{i_n}.\label{action1}
\eeq
On the other hand, the action $S[\Omega; \Lambda(\delta t)]$ in Eq.(\ref{expan2}) depend on the fields with the lower momentum, and the coupling constants which are defined at the lower cutoff $\Lambda(\delta t)$:
\beq
S[\Omega; \Lambda(\delta t)]=\sum_n \frac{1}{n!} \int_{p_1} \cdots \int_{p_n}\hat{\delta}(p_1+\cdots +p_n) g(\Lambda(\delta t)) \Omega_{i_1} \cdots \Omega_{i_n}.\label{action2}
\eeq
The difference between Eq.(\ref{action1}) and Eq.(\ref{action2}) lies in the variation of coupling constants when the cutoff is changed.
\beq
&&S[\Omega; \Lambda(\delta t)]-S[\Omega; \Lambda]|\nonumber\\
&=&\sum_n \frac{1}{n!} \int_{p_1}\cdots \int_{p_n} \hat{\delta}(p_1+\cdots +p_n) \left[ g(\Lambda(\delta t))-g(\Lambda) \right] \Omega_{i_1} \cdots \Omega_{i_n}\nonumber\\
&=&-\delta t \sum_n \frac{1}{n!} \int_{p_1}\cdots \int_{p_n} \hat{\delta}(p_1+\cdots +p_n) \left( \Lambda \frac{\partial}{\partial \Lambda} g(\Lambda) \right) \Omega_{i_1} (p_1) \cdots \Omega_{i_n} (p_n)\nonumber\\
&=&\frac{1}{2} \int_{p'} \tr \ln \left(\frac{\delta^2 S}{\delta \Omega^i \delta \Omega^j} \right)-\frac{1}{2}\int_{p'} \int_{q'} \frac{\delta S}{\delta \Omega^i} \left(\frac{\delta^2 S}{\delta \Omega^i \delta \Omega ^j} \right)^{-1} \frac{\delta S}{\delta \Omega^j}.\label{diff1}
\eeq

If the coupling constants depend on momenta, this cutoff dependence of the coupling constant is given by
\beq
\Lambda \frac{\partial}{\partial \Lambda}g(\Lambda)=\Lambda \frac{d}{d \Lambda}g(\Lambda)-\sum_i p^{\mu}_i \frac{\partial}{\partial p^{\mu}_i} g(\Lambda).
\eeq

In order to derive a differential equation for $S$, we transform all fields and coupling constants to dimensionless quantities. The mass dimension of a coupling constant $g$ is given
\beq
\dim[g]&=D-\sum_{\Omega_i} (d_{\Omega_i}+\gamma_{\Omega_i}),
\eeq
where $D$ denotes space-time dimensions, and $d_{\Omega_i}$, $\gamma_{\Omega_i}$ are the canonical and anomalous dimensions of $\Omega_i$.
Since the cutoff dependence of the coupling constant can be written
\beq
\Lambda \frac{d}{d \Lambda} g=\Lambda \frac{\partial}{\partial \Lambda}g -\dim[g]\cdot g+\sum_i p^{\mu}_i \frac{\partial}{\partial p^{\mu}_i}g(\Lambda),
\eeq
the WRG equation for the effective action $S$ is
\beq
\Lambda \frac{d}{d \Lambda}S&=&-\frac{1}{\delta t}\left[S[\Omega; \Lambda(\delta t)]-S[\Omega; \Lambda]| \right]\nonumber\\
&&-\left[D-\sum_{\Omega_i}\int_p \hat{\Omega}_i (p) \left(d_{\Omega_i} +\gamma_{\Omega_i} +\hat{p}^{\mu}\frac{\partial}{\partial \hat{p}^{\mu}} \right) \frac{\delta}{\delta \hat{\Omega_i}(p)} \right] \hat{S}\nonumber\\
&\equiv&-\frac{d}{dt}\hat{S},
\eeq
where the caret indicates dimensionless quantities.

Using Eq.(\ref{diff1}), we obtain WRG equation for dimensionless action:
\beq
\frac{d}{dt}S[\Omega; t]&=&\frac{1}{2\delta t} \int_{p'} \tr \ln \left(\frac{\delta^2 S}{\delta \Omega^i \delta \Omega^j}\right)\nonumber\\
&&-\frac{1}{2 \delta t}\int_{p'} \int_{q'} \frac{\delta S}{\delta \Omega^i (p')} \left(\frac{\delta^2 S}{\delta \Omega^i (p')\delta \Omega^j (q')} \right)^{-1} \frac{\delta S}{\delta \Omega^j (q')} \nonumber\\
&&+ \left[D-\sum_{\Omega_i} \int_p \hat{\Omega}_i (p) \left(d_{\Omega_i}+\gamma_{\Omega_i}+\hat{p}^{\mu} \frac{\partial}{\partial \hat{p}^{\mu}} \right) \frac{\delta}{\delta \hat{\Omega}_i (p)} \right] \hat{S}.\nonumber\\ \label{WRG-1} 
\eeq
The terms 
\beq
\int_{p'} \tr \ln \left(\frac{\delta^2 S}{\delta \Omega^i \delta \Omega^j}\right)
\eeq
and
\beq
\int_{p'} \int_{q'} \frac{\delta S}{\delta \Omega^i (p')} \left(\frac{\delta^2 S}{\delta \Omega^i (p')\delta \Omega^j (q')} \right)^{-1} \frac{\delta S}{\delta \Omega^j (q')}
\eeq
correspond to the one-loop and dumbbell diagrams respectively shown in Figures 1 and 2.

\begin{figure}[h]
\begin{center}
\includegraphics[width=4cm]{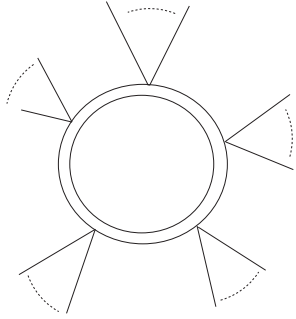}
\caption{one-loop diagram}
\end{center}
\end{figure}

\begin{figure}[h]
\begin{center}
\includegraphics[width=4cm]{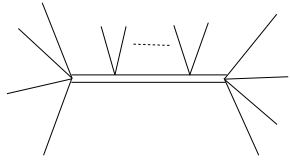}
\caption{dumbbell diagram}
\end{center}
\end{figure}

Here, the single and double lines represent the lower and higher modes, and the latter are integrated out in Eq.(\ref{WRG-1}). In the limit $\delta t \rightarrow 0$, the contribution of the higher order diagrams disappears.

If there are fermionic fields in addition to bosonic fields, the WRG equation (\ref{WRG-1}) can be rewritten:
\beq
\frac{d}{dt} S[\Omega;t]&=&\frac{1}{2\delta t} \int_{p'} str \ln \left(\frac{\delta^2 S}{\delta \Omega^i \delta \Omega^j}\right)\nonumber\\
&&-\frac{1}{2\delta t} \int _{p'} \int _{q'} (-1)^F \frac{\delta S}{\delta \Omega ^i (p')} 
\left(\frac{\delta^2 S}{\delta \Omega^i(p') \delta \Omega^j (q')}\right)^{-1} 
\frac{\delta S}{\delta \Omega^j(q')}\nonumber \\
&&+\left[D-\sum_{\Omega^i} \int_p \hat{\Omega}^i (p) 
(d_{\Omega^i} +\gamma_{\Omega^i} +\hat{p}^{\mu} 
\frac{\partial}{\partial \hat{p}^{\mu}}) 
\frac{\delta}{\delta \hat{\Omega}^i (p)}\right] \hat{S}\label{SWRG},
\end{eqnarray}
where 
\beq
str \ln \left(\frac{\delta^2 S}{\delta \Omega^i \delta \Omega^j}\right)&=&str \ln M_{i \bar{j}}\nonumber\\
&=&str \ln \left(
\begin{array}{cc}
M_{B B}& M_{B F}\\
M_{F B}& M_{F F}
\end{array}
\right)\nonumber\\
&=&\tr \ln M_{B B}-\tr \ln N_{F F},
\eeq
with
\beq
N_{FF}&=&M_{F F}-M_{F B} M^{-1}_{B B} M_{B F}. \label{str}
\eeq

The action $S[\Omega;t]$ in Eq.(\ref{WRG-1}) includes an infinite number of coupling constants $g_{i_1,\cdots,i_n}$, and Eq.(\ref{WRG-1}) gives infinite number of differential equations among them. To make these equations more tractable, we usually expand the effective action in powers of derivatives and maintain the first few terms. We often introduce a symmetry (supersymmetry, gauge symmetry, Z$_2$ symmetry etc) to further decrease the number of independent coupling constants.

Consider, for example, a single real scalar field theory which is invariant under $\varphi \rightarrow -\varphi$ (Z$_2$ symmetry). We can expand the most generic effective action as:
\beq
S[\varphi]=\int d^D x \Bigg[V[\varphi]+\frac{1}{2}K[\varphi] (\partial_\mu \varphi)^2 + H_1 [\varphi] (\partial_\mu \varphi)^4 +H_2 [\varphi](\partial_\mu \partial^\mu \varphi)^2 +\cdots \Bigg],\nonumber
\eeq
where $V[\varphi],K[\varphi],H_1[\varphi],\cdots$ are functions of $\varphi$. To second order of derivatives, this action becomes 
\beq
S[\varphi]=\int d^D x \Bigg[V[\varphi]+\frac{1}{2}K[\varphi](\partial_\mu \varphi)^2 \Bigg].
\eeq
Substituting this $S[\varphi]$ into Eq.(\ref{WRG-1}) and expanding the right hand side of Eq.(\ref{WRG-1}) up to $O(\partial^2)$, we obtain two differential equations for $V[\varphi],K[\varphi]$.

In scalar field theories, one often maintains only the local potential term $V[\varphi]$.
In this paper, we consider ${\cal N}=2$ SNL$\sigma$M and work at the second order of derivatives for the scalar fields.
In these theories, the local potential term is forbidden by the reparametrization invariance of the target manifold.

\section{Supersymmetric Nonlinear Sigma Model}
For ${\cal N}=2$ supersymmetric nonlinear sigma model (SNL$\sigma$M) in two dimensions (D=2), the action is determined by the K\"{a}hler potential $K[\Phi, \Phi^\dag]$:
\beq
S&=&\int dV K[\Phi,\Phi^\dag],\label{kahler}
\eeq
where
\beq
\int dV &\equiv& \int d^2 x d^2 \theta d^2 \bar{\theta}.\nonumber
\eeq
This is also true for ${\cal N}=2$ SNL$\sigma$M in D=3 and ${\cal N}=1$ SNL$\sigma$M in D=4, then we replace the space-time dimensions 2 with $D$.
Here, $\Phi$ is a chiral superfield and can be written by component fields:
\beq
\Phi^i(y)&=&\varphi^i (y)+\sqrt{2} \theta \psi^i(y)+\theta \theta F^i (y)\nonumber\\
&=&\varphi^i (x)+i \theta \sigma^\mu \bar{\theta} \partial_\mu \varphi^i (x) +\frac{1}{4} \theta \theta \bar{\theta} \bar{\theta} \partial^\mu \partial_\mu \varphi^i (x)\nonumber\\
&&+\sqrt{2} \theta \psi^i(x) -\frac{i}{\sqrt{2}}\theta \theta \partial_\mu \psi^i (x) \sigma^\mu \bar{\theta} +\theta \theta F^i(x)\\
&\equiv&\varphi^i(x)+\delta \Phi^i(x), \label{Phi-1}
\eeq
where,
\beq
y^\mu&=&x^\mu +i\theta \sigma^\mu \bar{\theta}  
\eeq
Using (\ref{Phi-1}), we expand the action (\ref{kahler}) around scalar fields ($\varphi,\varphi^*$). It is sufficient to expand it to $O((\delta \Phi)^4)$, because $\delta \Phi$ (and $\delta \Phi^\dag$) contains at least one $\theta$. Integrating out over Grassmann numbers ($\theta$ and $\bar{\theta}$), we obtain the SNL$\sigma$M action written by component fields:
\beq
S&=&\int d^D x \Bigg[g_{n \bar{m}}\left(\partial^{\mu} \varphi^n \partial_{\mu} \varphi^{* \bar{m}} +\frac{i}{2} \bar{\psi}^{\bar{m}} \sigma^{\mu}(D_{\mu} \psi)^n +\frac{i}{2} \psi^{n} \bar{\sigma}^{\mu}(D_{\mu} \bar{\psi})^{\bar{m}} +\bar{F}^{\bar{m}} F^{n}\right) \nonumber\\
&&-\frac{1}{2} K_{,nm \bar{l}} \bar{F}^{\bar{l}} \psi^n \psi^m -\frac{1}{2} K_{,n \bar{m} \bar{l}} F^{n} \bar{\psi}^{\bar{m}} \bar{\psi}^{\bar{l}}+\frac{1}{4} K_{,nm \bar{k} \bar{l}} (\bar{\psi}^{\bar{k}} \bar{\psi}^{\bar{l}})(\psi^n \psi^m)\Bigg].\label{action}
\eeq

First, we consider the bosonic part of the action (\ref{action}).
The loop correction term in the WRG eq.(\ref{WRG-1}) cannot be written in covariant form in general.
We use the K\"{a}hler normal coordinates (KNC) expansion of the action to obtain the covariant expression for the loop correction \cite{HN} \cite{HIN}.

The K\"{a}hler normal coordinates ($\omega,\omega^*$) are defined by the following condition:
\beq
K_{,\bar{j} i_1 \cdots i_n}(\omega,\omega^*)|_0 =0.\label{KNC-cond}
\eeq
Here, the index "0" denotes a value evaluated at the origin of KNC, which is the expansion point.

The Riemann normal coordinates (RNC) are well known in perturbative calculations \cite{AFM} \cite{BB}.
All geodesics in RNC become straight lines.
The RNC in K\"{a}hler manifolds, however, is not chiral and the coordinate transformation from the holomorphic coordinates to RNC is not holomorphic.
On the contrary, in KNC, geodesics cannot become straight lines, but the KNC preserves holomorphy.

Let us decompose arbitrary holomorphic coordinates ($z^i$) into a background field $\varphi^i$ and small fluctuation $\pi^i$ around it: $z^i=\varphi^i +\pi$.
The coordinate transformation from coordinates $z^i$ to KNC $\omega$ is given by
\beq
\omega^i&=&\pi^i +\sum^{\infty}_{n=2} \frac{1}{n!} [g^{i \bar{j}} K_{,\bar{j} i_1 \cdots i_n}(z^i,z^{* \bar{i}})]_{\varphi} \pi^{i_1} \cdots \pi^{i_n} \label{KNC}.
\eeq
These KNC $\omega^i$ is known to be transformed as the holomorphic tangent vectors.
Since they have the well-defined transformation property, we can use convenient coordinate system for our calculation.

In the D dimensional sigma model action
\beq
S_{scalar}=\int d^D x g_{i \bar{j}} (\varphi,\varphi^*) \partial^{\mu} \varphi^i \partial_{\mu} \varphi^{*\bar{j}},
\eeq
the background fields $\varphi^i$ are considered as the lower frequency modes and the fluctuation $\pi^i$ as the higher frequency modes.
The Euclidean path integral $Z$ in Eq.(\ref{expan1}) is written
\beq
Z=\int[D\varphi] [D\varphi^*] [D\pi] [D\pi] \exp \Big[-S[\varphi+\pi,\varphi^* +\pi] \Big].
\eeq
We regard the background field $\varphi^i$ as the origin of KNC, and transform the fluctuations $\pi^i$ to KNC fields $\bar{\pi}^i$:
\beq
Z \stackrel{KNC}{\longrightarrow} \int [D\varphi] [D\varphi^*] [D\bar{\pi}] [D\bar{\pi}^*] \exp \Big[-\bar{S}[\varphi+\bar{\pi},\varphi^*+\bar{\pi}^*] \Big],
\eeq
where bars indicate quantities in KNC, and $\bar{S}$ is given by
\beq
\bar{S}[\varphi+\bar{\pi},\varphi^*+\bar{\pi}^*]=\int d^D x \bar{g}_{i \bar{j}} (\varphi+\bar{\pi},\varphi^*+\bar{\pi}^*) \partial_{\mu}(\varphi+\bar{\pi})^i \partial^{\mu} (\varphi^*+\bar{\pi})^{\bar{j}}.\label{KNC-action}
\eeq
The KNC fields $\bar{\pi}(x)$ can be expanded by tangent vectors $\bar{\omega}(x)$, by solving Eq.(\ref{KNC}) for $\pi^i$ and taking KNC system
\beq
\bar{\pi}(x)=\bar{\omega}^i (x) -\frac{1}{2} {\bar{\Gamma}^i}_{\hspace{0.1cm}k_1 k_2}|_{\varphi} \bar{\omega}^{k_1}(x) \bar{\omega}^{k_2}(x) +O(\bar{\omega}^3).\label{pi-bar}
\eeq
Substituting Eq.(\ref{pi-bar}) into Eq.(\ref{KNC-action}) and using the property of KNC (\ref{KNC-cond}), we obtain the expansion of the sigma model action to the second order of the fluctuations in KNC:
\beq
\bar{S}[\varphi+\bar{\pi},\varphi^*+\bar{\pi}^*]\hspace{-0.3cm}&=&\hspace{-0.3cm}\int d^D x \bar{g}_{i \bar{j}}|_{\varphi}\left(\partial_{\mu} \varphi^i \partial^{\mu} \varphi^{* \bar{j}}+\partial_{\mu} \varphi^i \partial^{\mu} \bar{\omega}^{*\bar{j}}+ \partial_{\mu} \bar{\omega}^i \partial^{\mu} \varphi^{*\bar{j}}+\partial_{\mu} \bar{\omega}^i \partial^{\mu} \bar{\omega}^{* \bar{j}} \right)\nonumber\\
&&\hspace{-0.3cm}+\bar{g}_{i \bar{j},k \bar{l}}|_{\varphi} \left(\partial_{\mu} \varphi^i \partial^{\mu} \varphi^{*\bar{j}} \bar{\omega}^k \bar{\omega}^{*\bar{l}}+\frac{1}{2} \partial_{\mu} \varphi^i \partial^{\mu} \varphi^k \bar{\omega}^{* \bar{j}} \bar{\omega}^{* \bar{l}} +\frac{1}{2} \partial_{\mu} \varphi^{* \bar{j}} \partial^{\mu} \varphi^{* \bar{l}} \bar{\omega}^i \bar{\omega}^k \right)\nonumber\\
&&\hspace{-0.3cm}+O(\bar{\omega}^3).
\eeq

From this expansion, we can read the matrix
\beq
\left(\frac{\delta^2 S}{\delta \Omega^i (p') \delta \Omega^j (q')} \right)=M_{i j}(q',p')
\eeq
in Eq.(\ref{WRG-1}), whose trace corresponds to the one loop correction term.
These matrix elements are in momentum space 
\beq
M (q',p')=\left(
\begin{array}{cc}
M_{k \bar{l}}& M_{\bar{k} \bar{l}}\\
M_{k l}& M_{\bar{k} l}
\end{array}
\right),\label{matrix}
\eeq
where  
\beq
M_{k \bar{l}}&\equiv&\frac{\delta^2 \bar{S}}{\delta \varphi^{*\bar{l}}(q') \delta \varphi^{k}(p')}\nonumber\\
&=&\int_{p',q',P} (-p' \cdot q') \bar{g}_{k \bar{l}} (P) \hat{\delta}(p'+q'+P)\nonumber\\
&&+\int_{p',q',p,q,P''} (-p\cdot q) \bar{g}_{i \bar{j},k \bar{l}}(P''+p'+q') \varphi^i(p) \varphi^{* \bar{j}} \hat{\delta} (p'+q'+p+q+P''),\label{1-1}\nonumber\\
\\
M_{\bar{k} \bar{l}}&=&\int_{p',q',p,q,P''} (-p \cdot q) \bar{g}_{i \bar{k}, j \bar{l}}(P''+p'+q') \varphi^i (p) \varphi^j (q) \hat{\delta}(p'+q'+p+q+P''),\label{1-2}\nonumber\\
\\
M_{k l}&=&\int_{p',q',p,q,P''} (-p \cdot q) \bar{g}_{k \bar{i}, l \bar{j}}(P''+p'+q') \varphi^{* \bar{i}} (p) \varphi^{*\bar{j}} (q) \hat{\delta}(p'+q'+p+q+P''),\label{2-1}\nonumber\\
\\
M_{\bar{k} l}
&=&\int_{p',q',P} (-p' \cdot q') \bar{g}_{l \bar{k}} (P) \hat{\delta}(p'+q'+P)\nonumber\\
&&+\int_{p',q',p,q,P''} (-p\cdot q) \bar{g}_{i \bar{j},l \bar{k}}(P''+p'+q') \varphi^i(p) \varphi^{* \bar{j}} \hat{\delta} (p'+q'+p+q+P'').\label{2-2}\nonumber\\
\eeq
We write the (1,1) element (\ref{1-1}) as
\beq
M_{k \bar{l}}\equiv \langle \bar{l};-q' |\hat{M}_1|k;p' \rangle ,\label{M-1}
\eeq
to define the operator $\hat{M_1}$.
Here, the states $|k; p\rangle$ are defined by
\beq
\langle k;p|\bar{l};q \rangle =\langle \bar{l} ;p|k;q \rangle =\hat{\delta}(q-p)\delta_{k \bar{l}},
\eeq
\beq
\int_p \sum_{k \bar{l}} \left[|k;p \rangle \delta^{\bar{l} k} \langle \bar{l};p |+|\bar{l};p \rangle \delta^{\bar{l} k} \langle k;p|   \right]=1.
\eeq
From Eq.(\ref{M-1}), the explicit form of the operator $\hat{M}_1$ is
\beq
(\hat{M}_1)_{k \bar{l}}=\hat{p} \bar{g}_{k \bar{l}} \hat{p} + \bar{g}_{i \bar{j},k \bar{l}} \partial_{\mu} \varphi^i \partial^{\mu} \varphi^{*\bar{j}}.
\eeq
Similarly, from the other elements of the matrix (\ref{1-2}),(\ref{2-1}),(\ref{2-2}) we can define the following operators:
\beq
(\hat{M}_2)_{\bar{k} \bar{l}}&=&\bar{g}_{i \bar{k},j \bar{l}} \partial_{\mu} \varphi^i \partial^{\mu} \varphi^j,\\
(\hat{M}_3)_{k l}&=&\bar{g}_{k \bar{i},l \bar{j}} \partial_{\mu} \varphi^{* \bar{i}} \partial^{\mu} \varphi^{*\bar{j}},\\
(\hat{M}_4)_{\bar{k} l}&=&\hat{p} \bar{g}_{\bar{k} l} \hat{p} + \bar{g}_{i \bar{j},\bar{k} l} \partial_{\mu} \varphi^i \partial^{\mu} \varphi^{*\bar{j}}.
\eeq

The one loop correction term to $O(\partial^2)$ can be decomposed as follows:
\beq
\int_{p'} \tr \ln  M_{i j} (-p',p')= \int_{p'} \tr \ln \hat{M}_{1i \bar{j}} +\int_{p'} \tr \ln \hat{M}_{4 i \bar{j}},
\eeq
because $\hat{M}_2$ and $\hat{M}_3$ contribute only to higher derivative terms.
Noting that $\bar{g}_{i \bar{j},k}$ and $\bar{g}_{i \bar{j}, \bar{k}}$ are zero in KNC, we can calculate this trace:
\beq
\int_{p'} \tr \ln  M_{i j} (-p',p')&=&\frac{2}{(2\pi)^D}(\delta t) \int d \Omega_D \int d^D x \tr \ln \bar{g}_{i \bar{j}}\nonumber\\
&&+\frac{2}{(2\pi)^D} (\delta t) \int d \Omega_D \int d^D x \bar{g}_{i \bar{j},k \bar{l}} \bar{g}^{k \bar{l}}\partial_{\mu} \varphi^i \partial^{\mu} \varphi^{* \bar{j}}.\nonumber\\ \label{trace}
\eeq

Now we have to discuss the contribution of the fermion part in the supersymmetric action (\ref{action}).
It turns out that there is no fermion contribution to the bosonic action except for the first term in (\ref{trace}) with the opposite sign.
Therefore, the first term in the right hand side of Eq.(\ref{trace}) cancels with the similar contribution from the fermionic part.
In KNC, the second term can be written as
\beq
-\frac{2}{(2\pi)^D} (\delta t) \int d \Omega_D \int d^D x \bar{R}_{i \bar{j}} \partial_{\mu} \varphi^i \partial^{\mu} \varphi^{* \bar{j}}.
\eeq
Finally, we transform this result from KNC to the original coordinates. 
Then, the first term of the WRG eq.(\ref{WRG-1}) is
\beq
\frac{1}{2 \delta t}\int_{p'} \tr \ln  M_{i j}=-\frac{1}{(2\pi)^D} \int d \Omega_D \int d^D x R_{i \bar{j}} \partial_{\mu} \varphi^i \partial^{\mu} \varphi^{* \bar{j}}.
\eeq
The second term of Eq.(2.17), the contribution of the dumbell diagram,  vanishes to the order $O(\partial^2)$ in the derivative expansion, because external lines carrying soft momenta in the derivative expansion cannot satisfy the energy-momentum conservation law with the hard shell-momenta carried by the internal double lines in the dumbell diagram.

Similar derivation can be applied to D=2 ${\cal N}=2$, D=3 ${\cal N}=2$ and D=4 ${\cal N}=1$ SNL$\sigma$M.
We obtain the WRG equations for scalar part action for these SNL$\sigma$M.
\begin{itemize}
	\item In the case of D=2 ${\cal N}=2$ SNL$\sigma$M
 
\beq
&&\frac{d}{dt}\int d^2 x g_{i \bar{j}} (\partial_\mu \varphi)^i (\partial^\mu \varphi^*)^{\bar{j}}\nonumber\\
&=&\int d^2 x \Bigg[-\frac{1}{2\pi} R_{i \bar{j}}-\gamma \Big[\varphi^k g_{i \bar{j},k}+\varphi^{*\bar{k}}g_{i \bar{j},\bar{k}}+2g_{i \bar{j}} \Big] \Bigg](\partial_\mu \varphi)^i (\partial^\mu \varphi^*)^{\bar{j}}.\label{WRG-boson}\nonumber\\
\eeq

\item In the case of D=3 ${\cal N}=2$ SNL$\sigma$M

\beq
&&\frac{d}{dt}\int d^3 x g_{i \bar{j}} (\partial_\mu \varphi)^i (\partial^\mu \varphi^*)^{\bar{j}}\nonumber\\
&=&\int d^3 x \Bigg[-\frac{1}{2\pi^2} R_{i \bar{j}}-\gamma \Big[\varphi^k g_{i \bar{j},k}+\varphi^{*\bar{k}}g_{i \bar{j},\bar{k}}+2g_{i \bar{j}} \Big]\nonumber\\
&&\hspace{1cm}-\frac{1}{2}\Big[\varphi^k g_{i \bar{j},k}+\varphi^{*\bar{k}}g_{i \bar{j},\bar{k}} \Big] \Bigg](\partial_\mu \varphi)^i (\partial^\mu \varphi^*)^{\bar{j}}.
\eeq

\item In the case of D=4 ${\cal N}=1$ SNL$\sigma$M

\beq
&&\frac{d}{dt}\int d^4 x g_{i \bar{j}} (\partial_\mu \varphi)^i (\partial^\mu \varphi^*)^{\bar{j}}\nonumber\\
&=&\int d^4 x \Bigg[-\frac{1}{8\pi^2} R_{i \bar{j}}-\gamma \Big[\varphi^k g_{i \bar{j},k}+\varphi^{*\bar{k}}g_{i \bar{j},\bar{k}}+2g_{i \bar{j}} \Big]\nonumber\\
&&\hspace{1cm}-\Big[\varphi^k g_{i \bar{j},k}+\varphi^{*\bar{k}}g_{i \bar{j},\bar{k}} \Big] \Bigg](\partial_\mu \varphi)^i (\partial^\mu \varphi^*)^{\bar{j}}.
\eeq

\end{itemize}

In two dimensions, for example, the $\beta$ function of the K\"{a}hler metric is 
\beq
\frac{d}{dt}g_{i \bar{j}}&=&-\frac{1}{2\pi} R_{i \bar{j}}-\gamma \Big[\varphi^k g_{i \bar{j},k}+\varphi^{*\bar{k}}g_{i \bar{j},\bar{k}}+2g_{i \bar{j}} \Big]\nonumber\\
&\equiv&-\beta(g_{i \bar{j}}).\label{beta-metric}
\eeq

We concentrated to the discussion of the WRG equation for the bosonic part of the action.
In supersymmetric theories, Eq.(\ref{beta-metric}) has to be the scalar part of the relation among superfields.
Other parts will have contributions both from bosonic and fermionic parts.
Here, we simply assume supersymmetry to derive supersymmetric relation.
In this paper, we used straight cut-off in momentum space to obtain the WRG equation.
We will need some modification or counter terms to maintain supersymmetry.
If we assume supersymmetry, the WRG equation for K\"{a}hler potential should be 
\beq
\frac{d}{dt} \int dV K[\Phi,\Phi^\dag]&=&\int dV \triangle K_1 [\Phi,\Phi^\dag] \nonumber\\
&&\hspace{-3cm}+ \left[2-\sum_{\Omega^i} \int_p \hat{\Omega}^i (p) 
(d_{\Omega^i} +\gamma_{\Omega^i} +\hat{p}^{\mu} 
\frac{\partial}{\partial \hat{p}^{\mu}}) 
\frac{\delta}{\delta \hat{\Omega}^i (p)}\right] \hat{S},\label{WRG-poten}\nonumber\\
\eeq
where $\Omega$ stands for $\varphi,\varphi^*,\psi,\bar{\psi},F,\bar{F}$, and $\triangle K_1$ is the one-loop correction:
\beq
\triangle K_1=\frac{1}{2\pi} \ln \det g_{k \bar{l}}[\Phi,\Phi^{\dag}].
\eeq
Expanding Eq.(\ref{WRG-poten}) around scalar fields, as in Eq.(\ref{action}), we can obtain the WRG equations for various terms in Eq.(\ref{action}). These equations are consistent with Eq.(\ref{WRG-boson}).

From now on, we consider only D=2 ${\cal N}=2$ SNL$\sigma$M.

\section{Generic Einstein K\"{a}hler Manifolds}
Using the WRG equation for SNL$\sigma$M action (\ref{WRG-boson}), we find that SNL$\sigma$M are asymptotically free, when target spaces are the compact Einstein K\"{a}hler manifolds.

The Einstein K\"{a}hler manifolds satisfy the following condition:
\beq
R_{i \bar{j}}=\frac{h}{a^2} g_{i \bar{j}},\label{EK-condition}
\eeq
where $a$ is the radius of the manifold, which is related to the coupling constant $\lambda$ by
\beq
\lambda \equiv \frac{1}{a}.
\eeq
If the manifold is Hermitian symmetric space (G/H), the positive constant $h$ in Eq.(\ref{EK-condition}) is the eigenvalue of the quadratic Casimir operator in adjoint representation of global symmetry G \cite{HKN}.
For the C$P^N=SU(N+1)/[SU(N) \times U(1)]$ model, its value is $N+1$. 
We will discuss this model later.

When the manifolds have the radius $a=\frac{1}{\lambda}$, the scalar part of SNL$\sigma$M Lagrangian can be represented in the following form:
\beq
{\cal L}_{scalar}&=&g_{i \bar{j}}(\varphi,\varphi^*) \partial_{\mu} \varphi^i \partial^{\mu} \varphi^{* \bar{j}} \nonumber\\
&\stackrel{\varphi,\varphi^* \approx 0}{\longrightarrow}& \frac{1}{\lambda^2} \delta_{i \bar{j}} \partial_{\mu} \varphi^i \partial^{\mu} \varphi{* \bar{j}}.\label{lag-1}
\eeq
Although we consider only the scalar part for simplicity, the other parts are consistent with it.
We rescale the scalar fields as follows:
\beq
\frac{1}{\lambda} \varphi&=&\tilde{\varphi}.\label{rescale-1}
\eeq
Then, the Lagrangian (\ref{lag-1}) have the normalized kinetic term:
\beq
{\cal L}_{scalar}=\tilde{g}_{i \bar{j}}(\lambda \tilde{\varphi},\lambda \tilde{\varphi}^*) \partial_{\mu} \tilde{\varphi}^i \partial^{\mu} \tilde{\varphi}^{* \bar{j}}.
\eeq
Here, 
\beq
\tilde{g}_{i \bar{j}}|_{\tilde{\varphi},\tilde{\varphi}^*=0}=\delta_{i \bar{j}}
\eeq
is the metric of the manifold of unit radius.
Rescaling the WRG eq.(\ref{WRG-boson}) and comparing the coefficient of $\partial_{\mu} \tilde{\varphi}^i \partial^{\mu} \tilde{\varphi}^{* \bar{j}}$, we have
\beq
\frac{\partial}{\partial t} \tilde{g}_{i \bar{j}} (\lambda \tilde{\varphi},\lambda \tilde{\varphi}^*)&=&-\frac{1}{2\pi} \tilde{R}_{i \bar{j}} -\gamma [\tilde{\varphi}^k \tilde{g}_{i \bar{j},k }+\tilde{\varphi}^{* \bar{k}} \tilde{g}_{i \bar{j},\bar{k}}+2 \tilde{g}_{i \bar{j}}],\nonumber
\eeq
where $\tilde{R}_{i \bar{j}}$ is rescaled Ricci tensor and can be written as
\beq
\tilde{R}_{i \bar{j}}=h\lambda^2 \tilde{g}_{i \bar{j}}
\eeq
using Einstein K\"{a}hler condition (\ref{EK-condition}).

Because only $\lambda$ depends on $t$, this differential equation can be rewritten:
\beq
\frac{\dot{\lambda}}{\lambda} \tilde{\varphi}^k \tilde{g}_{i \bar{j},k } +\frac{\dot{\lambda}}{\lambda} \tilde{\varphi}^{* \bar{k}} \tilde{g}_{i \bar{j},\bar{k}}&=&-\left(\frac{h \lambda^2}{2\pi}+2\gamma  \right) \tilde{g}_{i \bar{j}}-\gamma[\tilde{\varphi}^k \tilde{g}_{i \bar{j},k }+\tilde{\varphi}^{* \bar{k}} \tilde{g}_{i \bar{j},\bar{k}}].\nonumber\\. 
\eeq
Since the left-hand side vanishes for $\varphi,\varphi^* \approx 0$, the coefficient of $\tilde{g}_{i \bar{j}}$ must vanish on the right-hand side. Thus, we obtain the anomalous dimension of scalar fields (or chiral superfields):
\beq
\gamma=- \frac{h \lambda^2}{4\pi}.\label{gamma}
\eeq
Comparing the coefficient of $\tilde{\varphi}^k \tilde{g}_{i \bar{j},k }$ (or $\tilde{\varphi}^{* \bar{k}} \tilde{g}_{i \bar{j},\bar{k}}$), we also obtain the $\beta$ function of $\lambda$:
\beq
\beta(\lambda)&\equiv&-\frac{d \lambda}{dt}\nonumber\\
&=&-\frac{h}{4\pi}\lambda^3.\label{beta}
\eeq

This $\beta$ function shows this theory is asymptotically free. The effective coupling constant $\lambda$ becomes small at the high energy. Thus at UV region the radius $a \rightarrow \infty$ and the manifolds become flat.

The Gell-Mann-Low function based on perturbation theory is discussed in ref.\cite{Morozov}. In that paper, the $\beta$ functions are calculated using specific forms of K\"{a}hler potential for the four group of compact homogeneous symmetric K\"{a}hler manifolds. Our method does not need such potential. We found the theories with Einstein K\"{a}hler manifolds are asymptotically free in general.
To conclude this section, we give two simple examples of Einstein K\"{a}hler manifolds. The constant $h$ for the hermitian symmetric space is given by the eigenvalue of the quadratic Casimir operator in adjoint representation of G as is shown in Table 1.

	
	\begin{table}[h]
	
	\begin{center}
	\begin{tabular}{|c|c|c|}
	\hline
	G/H                              & Dimensions (complex)  &$h$\\
	\hline \hline
	$SU(N)/[SU(N-1) \otimes U(1)]=$C$P^{N-1}$   & N-1        & N\\
	$SU(N)/[SU(N-M) \otimes U(M)]$   & M(N-M)                & N\\
	$SO(N)/[SO(N-2) \otimes U(1)]=Q^{N-2}$   & N-2           & N-2\\
	$Sp(N)/U(N)$                     & $\frac{1}{2}N(N+1)$   & N+1\\
	$SO(2N)/U(N)$                    & $\frac{1}{2}N(N+1)$   & N-1\\
	$E_{6}/[SO(10) \otimes U(1)]$    &16                     &12\\
	$E_{7}/[E_6 \otimes U(1)]$        &27                     &18\\
	\hline
	\end{tabular}
	\caption{The values of $h$ for hermitian symmetric spaces}
	\end{center}
	\end{table}
	
	
	\begin{itemize}
	\item  C$P^N$ model

	The K\"{a}hler potential involves a set of N chiral and anti-chiral superfield $\vec{\Phi}=(\Phi^1,\cdots,\Phi^N)$
	\beq
	K[\Phi,\Phi^\dag]=a^2 \ln (1+\vec{\Phi} \vec{\Phi}^\dag).
	\eeq
	From this K\"{a}hler potential, we can obtain K\"{a}hler metric and Ricci tensor. 
	\beq
	g_{i \bar{j}}&\equiv&\partial_i \partial_{\bar{j}}K   \nonumber\\
	&=&a^2 \left(\frac{\delta_{i \bar{j}}}{1+\vec{\varphi} \vec{\varphi}^*} \right)\label{CP-metric}\\
	R_{i \bar{j}}&\equiv&-\partial_{\bar{j}} \partial_i (\ln \det g_{k \bar{l}})\nonumber\\
	&=&\frac{N+1}{a^2} g_{i \bar{j}}\label{CP-Ricci}.
	\eeq
	Eq.(\ref{CP-Ricci}) shows that this manifold is Einstein K\"{a}hler manifold with $h=N+1$. Hence Eqs.(\ref{gamma}) and (\ref{beta}) give us immediately 
	\beq
	\gamma&=&-\frac{(N+1)\lambda^2}{ 4 \pi}\\
	\beta(\lambda)&=&-\frac{(N+1)\lambda^3}{4 \pi}.
	\eeq 
	These results are consistent with the perturbative results \cite{Morozov}.

	\item $Q^N$ model

	The next example is $Q^N=SO(N+2)/[SO(N) \times U(1)]$ model, whose K\"{a}hler potential has the form
	\beq
	K[\Phi,\Phi^{\dag}]=a^2 \ln \left(1+\vec{\Phi} \vec{\Phi}^\dag +\frac{1}{4} \vec{\Phi}^2 \vec{\Phi}^{\dag 2} \right),
	\eeq
	where $\vec{\Phi}=(\Phi^1,\cdots,\Phi^N)$.
	K\"{a}hler metric and Ricci tensor are given by
	\beq
	g_{i \bar{j}}&=&a^2 \Bigg[\frac{\delta_{i \bar{j}}}{1+\vec{\varphi} \vec{\varphi}^* +\frac{1}{4} \vec{\varphi}^2 \vec{\varphi}^{* 2}}\nonumber\\
	&&+\frac{\varphi^i \varphi^{* \bar{j}} \left(1+\vec{\varphi} \vec{\varphi}^* \right)-\left(\varphi_i^* \varphi_{\bar{j}} +\frac{1}{2} \vec{\varphi}^2 \varphi_i^* \varphi^{* \bar{j}} +\frac{1}{2} \vec{\varphi}^{* 2} \varphi^i \varphi_{\bar{j}} \right) }{ \left(1+\vec{\varphi} \vec{\varphi}^* +\frac{1}{4} \vec{\varphi}^2 \vec{\varphi}^{* 2}  \right)^2 }  \Bigg], \label{Q-metric}\nonumber\\
	\\
	R_{i \bar{j}}&=&\frac{N}{a^2}g_{i \bar{j}} \label{Q-Ricci}.
	\eeq
	Eq.(\ref{Q-Ricci}) shows that this manifold is also Eistein K\"{a}hler manifold, and $h=N$.
	According to the same discussion as C$P^N$ model, we obtain the anomalous dimension and $\beta$ function:
	\beq
	\gamma&=&-\frac{N \lambda^2}{ 4 \pi}\\
	\beta(\lambda)&=&-\frac{N \lambda^3}{4 \pi}.
	\eeq

\end{itemize}

Combining these two examples, we will construct a new model in the next section.

\section{A New Model}
Consider the following K\"{a}hler potential which has two parameters $\lambda(t)=\frac{1}{a(t)}$ and $g(t)$:
\beq
K[\Phi,\Phi^{\dag}]=a^2 ( t) \ln \left(1+\vec{\Phi} \vec{\Phi}^{\dag} +g(t) \vec{\Phi}^2 \vec{\Phi}^{\dag 2} \right).
\eeq
When $g=0$, this is C$P^N$ model, and this becomes $Q^N$ model when $g=\frac{1}{4}$. The former has the global symmetry $SU(N+1)$, and the latter has $SO(N+2)$.

This K\"{a}hler potential gives the K\"{a}hler metric and Ricci tensor:
\beq
g_{i \bar{j}}&=&a^2 (t) \Bigg[\frac{\delta_{i \bar{j}}}{1+\vec{\varphi} \vec{\varphi}^* +g(t) \vec{\varphi}^2 \vec{\varphi}^{*2}}\nonumber\\
&&+\frac{4 g(t)\varphi^i \varphi^{*\bar{j}}(1+\vec{\varphi} \vec{\varphi}^*)-(\varphi_i^* \varphi_{\bar{j}} +2g(t)\vec{\varphi}^2 \varphi_i^* \varphi^{*\bar{j}}+2g(t) \vec{\varphi}^{*2}\varphi^i \varphi_{\bar{j}})}{(1+\vec{\varphi} \vec{\varphi}^* +g(t) \vec{\varphi}^2 \vec{\varphi}^{*2})^2}  \Bigg],\nonumber\\
R_{i \bar{j}}&=&(N+1) \frac{1}{a(t)} g_{i \bar{j}}-\Bigg[\frac{4g(t)\delta_{i \bar{j}}}{1+4g(t)\vec{\varphi} \vec{\varphi}^* +g(t) \vec{\varphi}^2 \vec{\varphi}^{*2}} \nonumber\\
&&+\frac{16g^2 (t) \varphi^i \varphi^{* \bar{j}} \vec{\varphi} \vec{\varphi}^{*} -16 g^2 (t) \varphi_i^* \varphi_{\bar{j}} -8g^2 (t) \left(\vec{\varphi}^2 \varphi_i^* \varphi^{*\bar{j}}+\vec{\varphi}^{*2}\varphi^i \varphi_{\bar{j}} \right)}{(1+4g(t)\vec{\varphi} \vec{\varphi}^* +g(t) \vec{\varphi}^2 \vec{\varphi}^{*2})^2} \Bigg].\label{new-Ricci}\nonumber\\
\eeq
Note that Eq.(\ref{new-Ricci}) shows this manifold is not an Einstein K\"{a}hler manifold unless $g$ takes specific values. 
According to the same discussion as section 4, we obtain
\beq
\gamma&=&-\frac{\lambda^2}{4\pi}[(N+1) -4g]\label{gamma-new},\\
\beta(\lambda)&=&-\frac{\lambda^3}{4\pi}[(N+1)+8 g(2g-1)], \label{beta-new1}\\
\beta(g)&=&\frac{4\lambda^2}{\pi}g^2 (4g-1)\label{beta-new2}.
\eeq
Fig.3 shows the renormalization group flows in $\lambda-g$ plane.

\begin{figure}
\begin{center}
\psfrag{g}{\Large$g$}
\psfrag{lambda}{\Large$\lambda$}
\includegraphics[width=7cm]{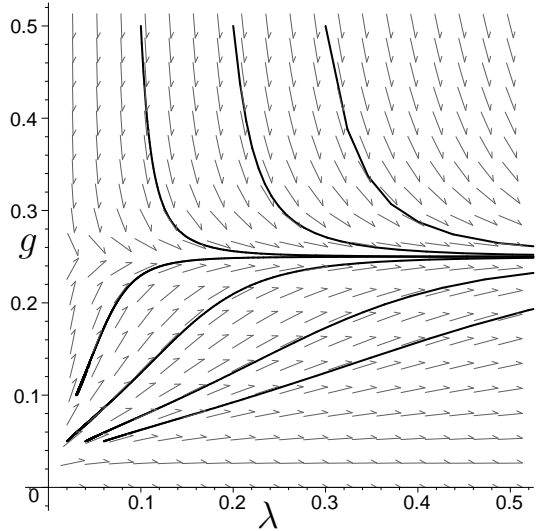}
\caption{Renormalization group flows (The direction of arrows shows infrared region.)}
\label{fig:flow.eps}
\end{center}
\end{figure}
From Eq.(\ref{beta-new2}), $\beta(g)$ is zero when $g=0$ and $g=\frac{1}{4}$.
These values of $g$ correspond to C$P^N$ and $Q^N$ models respectively, and in both cases the manifolds become Einstein K\"{a}hler manifolds. 

In the infrared(IR) region, the coupling constant $\lambda$ becomes infinity and $g$ approaches $\frac{1}{4}$.
Thus this model becomes $Q^N$ model with strong coupling at low energy.
On the contrary, at ultraviolet(UV) region, the flow separates $g<\frac{1}{4}$ and $g>\frac{1}{4}$ regions.
In $g<\frac{1}{4}$ region, the coupling constant $\lambda \rightarrow 0$ and $g \rightarrow 0$.
Thus C$P^N$ with the radius $a \rightarrow \infty$ is the UV fixed point of this new model.
In $g>\frac{1}{4}$ region, the coupling constant $g$ goes to infinity and $\lambda$ becomes zero.

\section{Conclusions}
In this paper we have found the $\beta$ function for the WRG equation in two dimensional ${\cal N}=2$ supersymmetric nonlinear sigma model using nonperturbative method. 
This $\beta$ function of K\"{a}hler metric at $O(\partial^2)$ approximation consists of the one-loop correction term and rescaling term. 
We have obtained several new results from this $\beta$ function.

First, we have seen all theories are asymptotically free if target spaces are generic compact Einstein K\"{a}hler manifolds.
This statement is drawn by using only Einstein K\"{a}hler condition, and does not depend on the specific form of K\"{a}hler potential.

Second, we have constructed a new model, which has two parameters $g$ and $\lambda$.
This model is not Einstein K\"{a}hler manifold except for special values of $g$.
Using the Wilsonian renormalization group equation, we obtain two $\beta$ functions of each parameters.
One of the $\beta$ functions ($\beta(g)$) is zero when the model becomes C$P^N$ and $Q^N$ model. 
These two models have the different global symmetry $SU(N+1)$ and $SO(N+2)$ respectively, and both of them are Einstein K\"{a}hler manifolds.
The renormalization group flow shows the new model interpolates C$P^N$ model and $Q^N$ model.
We have also found the C$P^N$ model is the UV fixed point of the model.
From this analysis, we conjecture that generic K\"{a}hler manifolds go to Einstein K\"{a}hler manifolds through Wilsonian renormalization group flow.

\section*{Acknowledgements}
We would like to thank Muneto Nitta for useful communication in early stage of this work.


\end{document}